\documentclass[AMA,STIX2COL]{MRM}
\articletype{Submitted to Magnetic Resonance in Medicine}%

\usepackage{pdfpages}

\received{\today}
\revised{}
\accepted{}
\begin{document}

\title{Contrast-Optimized Basis Functions for Self-Navigated Motion Correction in Quantitative MRI}

\author[1,2]{Elisa Marchetto*}{\orcid{0000-0001-7904-8434}}

\author[1,2]{Sebastian Flassbeck*}{\orcid{0000-0003-0865-9021}}

\author[1,2,3]{Andrew Mao}{\orcid{0000-0002-1398-0699}}

\author[1,2]{Jakob Assl{\"{a}}nder}{\orcid{0000-0003-2288-038X}}

\authormark{Marchetto \& Flassbeck \textsc{et al.}}

\address[1]{\orgdiv{Center for Biomedical Imaging, Dept. of Radiology}, \orgname{NYU School of Medicine}, \orgaddress{\state{NY}, \country{USA}}}
\address[2]{\orgdiv{Center for Advanced Imaging Innovation and Research (CAI\textsuperscript{2}R), Dept. of Radiology}, \orgname{NYU School of Medicine}, \orgaddress{\state{NY}, \country{USA}}}
\address[3]{Vilcek Institute of Graduate Biomedical Sciences, New York University Grossman School of Medicine, New York, New York, USA.}

\corres{Corresponding author: Jakob Assl{\"{a}}nder. \email{jakob.asslaender@nyulangone.org}}

\presentaddress{Center for Biomedical Imaging, NYU School of Medicine, 227 E 30$^{th}$ Street, New York, NY 10016, USA.}

\finfo{This work was supported by \fundingAgency{NIH/NINDS} grants \fundingNumber{R01~NS131948}, \fundingNumber{F30~AG077794} and \fundingAgency{NIH/NIBIB} grant \fundingNumber{P41~EB017183}.}

\abstract[Summary]{
\textbf{Purpose:} The long scan times of quantitative MRI techniques make them vulnerable to motion artifacts. For MR-Fingerprinting-like approaches, this problem can be addressed with self-navigated retrospective motion correction based on reconstructions in a singular value decomposition (SVD) subspace. However, the SVD promotes high signal intensity in all tissues, which limits the contrast between tissue types and ultimately reduces the accuracy of registration. The purpose of this paper is to rotate the subspace for maximum contrast between two types of tissue and improve the accuracy of motion estimates.

\textbf{Methods:} A subspace is derived that promotes contrasts between brain parenchyma and CSF, achieved through the generalized eigendecomposition of mean autocorrelation matrices, followed by a Gram-Schmidt process to maintain orthogonality. We tested our motion correction method on 85 scans with varying motion levels, acquired with a 3D hybrid-state sequence optimized for quantitative magnetization transfer imaging.

\textbf{Results:} A comparative analysis shows that the contrast-optimized basis significantly improves the parenchyma-CSF contrast, leading to more accurate motion estimates and reduced artifacts in the quantitative maps.

\textbf{Conclusion:} The proposed contrast-optimized subspace improves the accuracy of the motion estimation. 
}

\wordcount{3500/5000 - Figure count: 7/10}

\jnlcitation{\cname{% 
\author{E. Marchetto},
\author{S. Flassbeck},
et al. (\cyear{2024}), 
\ctitle{Contrast-Optimized Basis Functions for Self-Navigated Motion Correction in 3D Quantitative MRI}, \cjournal{Magn. Reson. Med.}, \cvol{xxxx;xx:x--x}.}}

\maketitle
\footnotetext{*\,E. Marchetto and S. Flassback contributed equally to this work.}

\section{Introduction}\label{sec1}
Magnetic resonance fingerprinting (MRF) is a multiparametric quantitative magnetic resonance imaging (qMRI) approach. Its key concept is a variation of sequence parameters between repetition times (TRs) to maintain a transient state of the magnetization. \cite{Ma2013, panda2017magnetic}

Motion-induced artifacts pose significant challenges in MRF-like experiments, whether due to physiological motion or involuntary motion due to pathological conditions.
3D acquisitions mitigate the risk of through-plane motion\cite{yu2018exploring}---which is prohibitively difficult to correct retrospectively---but the associated long scan times make motion more likely.

Both real-time and retrospective motion correction strategies rely on navigators or external tracking devices to provide estimates of the subject's motion.\cite{zaitsev2015motion}
Fat-navigators have previously been proposed for robust retrospective motion correction in structural brain MRI,\cite{gallichan2016retrospective, marchetto2023robust, marchetto2024analysis} and they have successfully been implemented in a 3D MRF sequence.\cite{hu2023improving} However, navigator-based motion correction techniques depend on the sequence timing, as they are typically acquired during waiting periods. Moreover, even fat-selective excitation pulses can perturb the spin dynamics via the magnetization transfer (MT) effect, which can compromise quantitative measurements.

On the other hand, external tracking devices are, in general, faster than navigator-based techniques and monitor the subject's movements independently from the acquired sequence. Such systems include wearable sensors and optical tracking devices.\cite{madore2023external} However, despite being sequence agnostic, external tracking devices have seen limited use in the MRF community,\cite{kretzler2025free} likely due to the challenges associated with their clinical implementation, e.g., complexity of system setup.\cite{maclaren2022external}

An alternative approach is self-navigated motion correction, which requires no additional data as the motion parameter estimation is derived from the acquired data itself. The Periodically Rotated Overlapping Parallel Lines with Enhanced Reconstruction (PROPELLER) is a widely used self-navigated imaging technique in which consecutively rotated blades are acquired around the k-space center.\cite{pipe1999motion} Each blade can be reconstructed into a low-resolution image, which is then used to detect and correct motion artifacts. PROPELLER-like acquisition schemes require long acquisition times and are not well suited for dynamic imaging protocols in which flip angles and pulse durations change with every TR.

Self-encoded FID navigators have also shown compelling results for self-navigated motion correction using a 3D radial trajectory.\cite{wallace2024rapid} However, they necessitate the acquisition of prescan normalized data prior to the scan to calibrate the motion model. Additionally, the associated spin perturbations would likely need to be accounted for in the model when combining FID navigators with quantitative MRI.

In 3D MRF acquisitions, a self-navigated motion correction approach was previously proposed. \cite{kurzawski2020retrospective}
Kurzawski et al. reconstructed brain navigators from 7\,s segments in a sub-space spanned by the truncated singular value decomposition (SVD) subspace. This method leverages the low-rank nature of the underlying data to produce coefficient images, which are then used to extract the motion estimates utilizing rigid registration. However, the SVD promotes high signal intensity in all types of tissue, which limits the contrast and can ultimately reduce the accuracy of the registration and extracted motion estimates.

In this work, we aim to improve the accuracy of motion estimates by deriving a basis that maximizes the signal of the fingerprints corresponding to brain parenchyma (i.e., white and gray matter), while minimizing the signal of fingerprints from cerebrospinal fluid (CSF), therefore actively promoting the contrast-to-noise ratio.

\section{Theory}\label{theory}\label{sec21}
The proposed contrast-optimized basis was inspired by the concept of region-optimized virtual (ROVir) coils, where the generalized eigendecomposition was used to maximize the signal-to-interference ratio, yielding promising results in suppressing unwanted signals in various MRI applications.\cite{kim2021region} We propose to use the generalized eigendecomposition to maximize the contrast-to-noise ratio between two types of tissue. This approach effectively rotates the SVD subspace, resulting in a \textit{contrast-optimized} basis that promotes the contrast in the first and last coefficient images.

In the following, we will translate the ROVir formalism to subspace modeling.
While this approach can be applied to any two types of tissue, we will outline the concept in the example of brain parenchyma and CSF, and we will use simulated fingerprints to calculate the basis. 
First, we calculate an SVD and truncate it heuristically to the first 3 basis functions. This step ensures that the subspace encompasses most of the signal intensity, which minimizes artifacts from unmodeled signals.

%% Basis figure 1
\begin{figure}[htbp]
    \centering
    \includegraphics[width=0.95\linewidth]{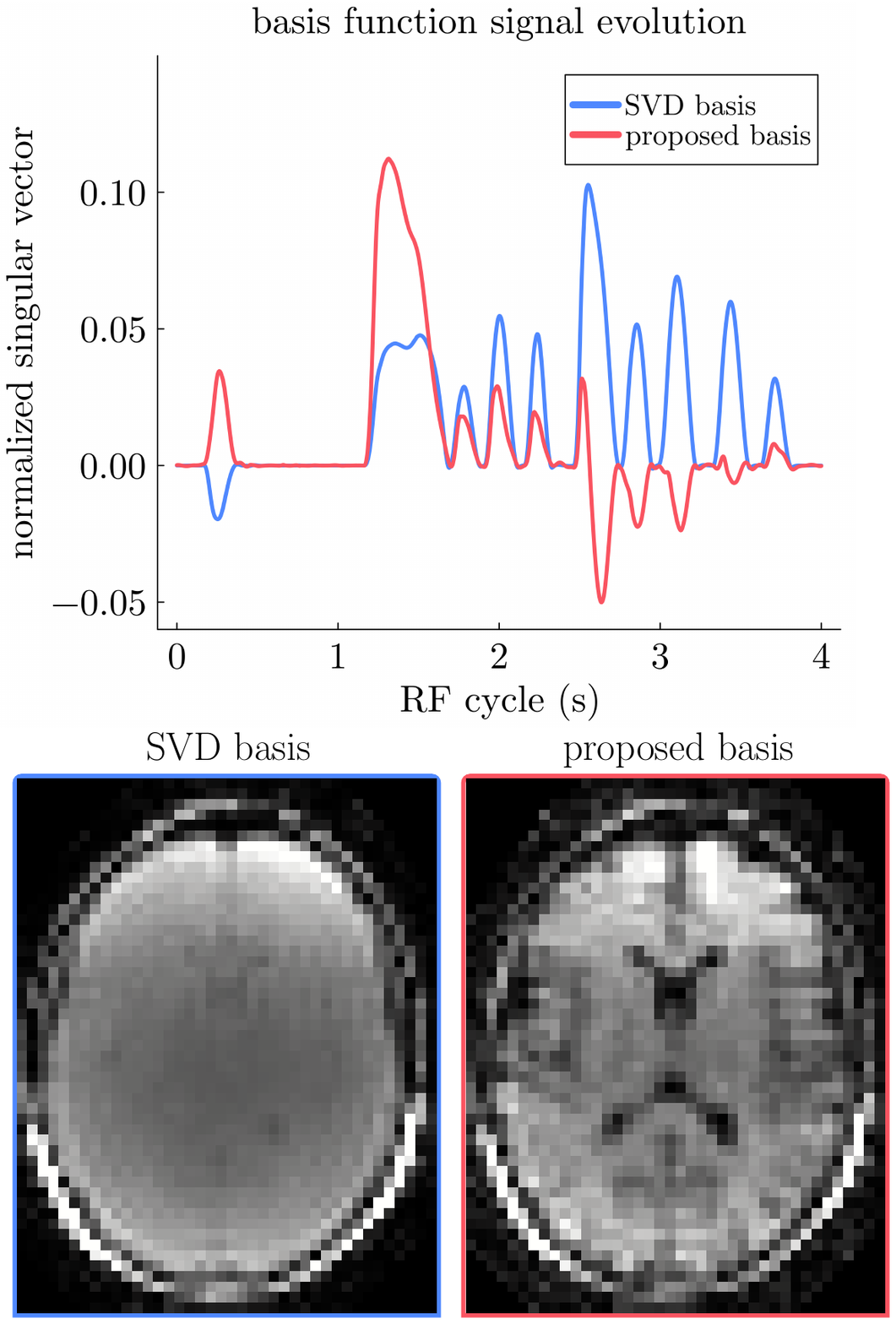}
    \vspace{-10pt}
    \caption{Top: Exemplary basis functions $\mathbf{U}_{\text{SVD}}^{(1)}$ and $\mathbf{U}_{\text{opt}}^{(1)}$, which are used to reconstruct low-resolution images for motion estimation (bottom), derived by aggregating the radial spokes acquired during one 4\,s RF cycle (cf. Sec.\,\ref{sec32}). 
    The SVD basis maximizes the signal across all tissues, discouraging contrast between them.
    In contrast, the proposed basis directly maximizes the contrast between tissues (in this case, brain parenchyma and CSF). 
    We note that the contrast in the SVD basis exhibits considerable variability between sequences, choices of parameters used for simulating each tissue, etc., since contrast is not part of the optimization objective.
    Here, we show a case with particularly low contrast, more examples can be found in Supplementary Fig.\,S1.}
    \label{fig:basisplot}
\end{figure}

In the second step, we project the fingerprints for brain parenchyma $\mathbf{s}_{\text{b}}$ and CSF $\mathbf{s}_{\text{f}}$ into the SVD subspace:
\begin{equation}
    \newcommand{\dd}[1]{\mathrm{d}#1}
    \mathbf{c}_{\text{b}} = \mathbf{U}_{\text{SVD}} \mathbf{s}_{\text{b}}
    \label{eq:usbrain}
\end{equation}
\begin{equation}
    \newcommand{\dd}[1]{\mathrm{d}#1}
    \mathbf{c}_{\text{f}} = \mathbf{U}_{\text{SVD}} \mathbf{s}_{\text{f}}
    \label{eq:uscsf}
\end{equation}
and calculate the mean autocorrelation matrices $\mathbf{C}_{\text{b}}$ and $\mathbf{C}_{\text{f}}$:
\begin{equation}
    \mathbf{C}_{\text{b}} = \mathbf{c}_{\text{b}}^H \mathbf{c}_{\text{b}}
    \label{eq:A}
\end{equation}
\begin{equation}
    \mathbf{C}_{\text{f}} = \mathbf{c}_{\text{f}}^H \mathbf{c}_{\text{f}}.
    \label{eq:B}
\end{equation}
The subspace that optimizes the contrast between brain parenchyma and CSF is given by the weights $\mathbf{w}$ that maximize 
\begin{equation}
    \mathbf{U_{\text{opt}}} \triangleq \frac{\mathbf{w}^H\mathbf{C}_{\text{b}}\mathbf{w}}{\mathbf{w}^H\mathbf{C}_{\text{f}}\mathbf{w}},
    \label{eq:SIR}
\end{equation}
thus, maximizing the signal fingerprints of the brain parenchyma while minimizing the signal fingerprints of the CSF.
Since $\mathbf{C}_{\text{b}}$ and $\mathbf{C}_{\text{f}}$ are positive-semidefinite Hermitian-symmetric matrices, and $\mathbf{C}_{\text{f}}$ has full rank, there is a set of eigenvalues $\mathbf{\lambda}_i$ with linearly independent eigenvectors $\mathbf{w}_i$ such that
\begin{equation}
    \mathbf{C}_{\text{b}}\mathbf{w}_i = \mathbf{\lambda}_i \mathbf{C}_{\text{f}}\mathbf{w}_i.
    \label{eq:eigen}
\end{equation}
Here, $i \in \{1,2,3\}$.
Eq.~\eqref{eq:eigen} can be solved by calculating the generalized eigendecomposition \cite{golub2013matrix} for $\mathbf{C}_{\text{b}}$ and $\mathbf{C}_{\text{f}}$ and order the generalized eigenvalues and eigenvectors so that $\mathbf{\lambda}_1\geq\mathbf{\lambda}_2\geq\mathbf{\lambda}_3$. Assuming normalized eigenvectors ($||\mathbf{w}_i||_2=1$), we can rotate the basis functions with 
\begin{equation}
    \mathbf{U}_{\text{opt}}^{(i)} = \mathbf{U}_{\text{SVD}}^{(i)} \mathbf{w}_i.
    \label{eq:gdecomp}
\end{equation}
The subspace maximizes the parenchyma and minimizes the CSF signal in the first coefficient, while the last coefficient has the opposite properties. The Gram-Schmidt approach is then applied to ensure the orthogonality between the three bases, leaving the first basis function unchanged. The resulting brain basis function $\mathbf{U}_{\text{opt}}^{(1)}$ and coefficient image are shown in Fig.~\ref{fig:basisplot}, which highlights the improved contrast between parenchyma and CSF compared to the SVD basis $\mathbf{U}_{\text{SVD}}^{(1)}$.

\section{Methods}\label{methods}\label{sec3}
\subsection{Data Acquisition} \label{sec31}
The acquired data comprised 11 healthy control scans and 75 scans from participants affected by mild Traumatic Brain Injury, for a total of 86 acquisitions. The participants were instructed to stay still during the scan. We used a 3\,T Prisma scanner (Siemens Healthineers, Erlangen, Germany) on which we ran a 3D hybrid-state sequence\cite{asslander2019hybrid} optimized for quantitative magnetization transfer (qMT) imaging.\cite{asslander2024unconstrained} Each RF pulse pattern is 4\,s long and consists of an inversion pulse $\pi$, followed by a train of 1141 rectangular RF pulses with a varying flip-angle and pulse duration and spaced 3.5\,ms apart. 

Based on Henkelman's two-pools model,\cite{henkelman1993quantitative} the free-water pool (denoted by the superscript $f$) exchanges magnetization with the semi-solid spin pool (denoted by the superscript $s$) at the rate $R_x$. Each pool has a fractional pool size, denoted as $m_0^f$ and $m_0^s$, where $m_0^f+m_0^s=1$. Additionally, each pool’s spin dynamics is captured by respective longitudinal ($R_1^f$, $R_1^s$) and transverse ($R_2^f$, $T_2^s$) relaxation rates, which are the inverse of respective relaxation times. We used six different flip-angle patterns, optimized to encode these six core biophysical MT parameters: $m_0^s$, $R_1^f$, $R_2^f$, $R_\text{x}$, $R_1^s$, and $T_2^s$ (note that we use the relaxation time $T_2^s$ for historic reasons).\cite{asslander2024unconstrained}

The sequence utilizes a 3D radial koosh-ball readout trajectory with reordered golden-angle increments \cite{Winkelmann2007, Ehses2013,feng2014golden,flassbeck2024eddycurrent} and nominal resolution of 1\,mm isotropic ($|k_{max}|=\pi/1\,mm$). Each RF pattern is repeated 30 times, for a total scan time duration of 12\,min. For each subject, we also acquired a 3D MP-RAGE with 1\,mm isotropic resolution. Informed consent was obtained prior to the scan in accordance with our Institutional Review Board.

%% Figure 2 - simulated motion pars
\begin{figure*}[htb]
    \centering
    \includegraphics[width=0.9\linewidth]{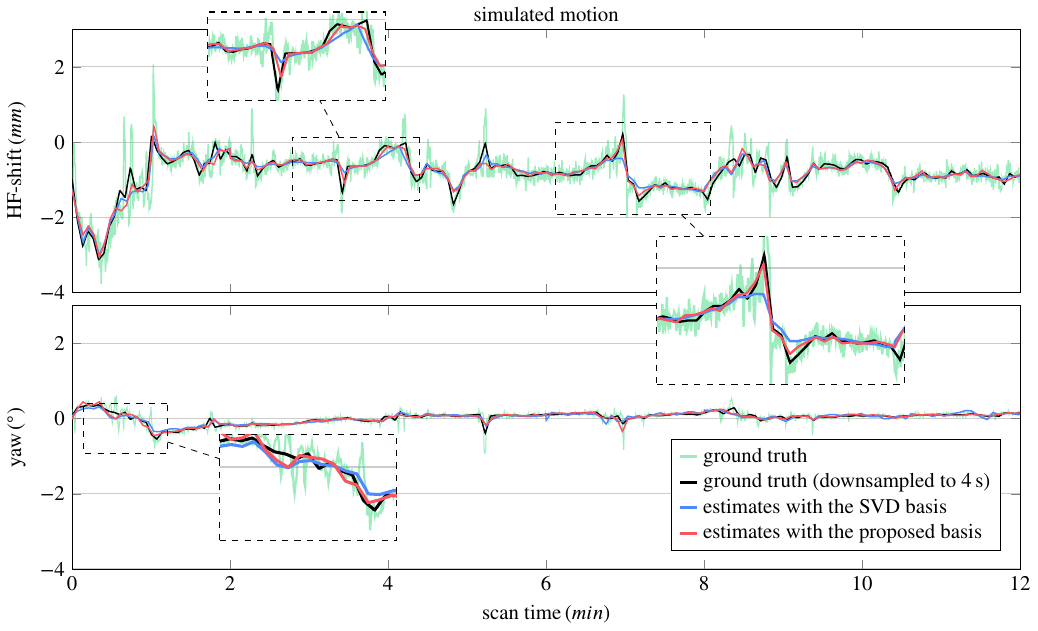}
    \vspace{-10pt}
    \caption{Estimates of simulated motion, using the SVD and proposed basis. The ground truth motion was obtained from an online database and has a motion score of 1.77\,mm. The magnifications highlight improvements of the proposed over the SVD basis, which is corroborated by a reduction in the RMSE score from 0.38\,mm to 0.35\,mm (lower bound of 0.3\,mm calculated between the ground truth and the downsampled ground truth). Here, we show one representative translation (head-foot) and one representative rotation. All 6 motion parameters can be found in the Supporting Fig.\,S4.}
    \label{fig:mopar_simu}
\end{figure*}

\subsection{Motion Estimation and Image Reconstruction}\label{sec32}
We aggregated all acquired spokes from each 4\,s RF cycle to reconstruct low-resolution coefficient images (4\,mm isotropic) in the subspace of the SVD and the proposed contrast-optimized basis.\cite{Tamir2017} The reconstruction problem can be formulated as: 
\begin{equation}
    \newcommand{\dd}[1]{\mathrm{d}#1}
    \hat{x} = \text{argmin}_x \frac{1}{2} \mid\mid{Ex-y}\mid\mid ^2_2 + \lambda_t\mid\mid{\nabla_{t}x}\mid\mid_1\
    \label{eq:model1}
\end{equation}
where \(E\) represents the encoding operator, which combines gridding, Fourier transform, and coil sensitivities. We used a total variation (TV) penalty along time to mitigate undersampling artifacts and noise,\cite{feng2014golden,block2007undersampled} where the associated regularization strength $\lambda$ was chosen based on the simulations described in Sec.\,\ref{sec333}. The finite difference operator is denoted by \(\nabla_{t}\).
The reconstruction problem was solved using the Alternating Directions Method of Multipliers solver.\cite{boyd2011distributed} Examples of low-resolution coefficient images reconstructed either using the SVD basis or the proposed contrast-optimized basis for each of the six flip-angle patterns are shown in Fig.~\ref{fig:basisplot} and Supporting Fig.\,S1.

The $N=180$ reconstructed low-resolution volumes were rigidly co-registered separately for each flip-angle pattern with the \textit{Statistical Parametric Mapping} (SPM) software.\cite{SPM12} More details regarding the extraction of the motion parameters can be found in Supporting Sec.\,S1.
The resulting affine matrices were then used to rotate the k-space trajectory of each 4\,s block. Translations were incorporated by multiplying the k-space data with a corresponding linear phase slope. 

Due to the limited 4\,s temporal resolution of the motion correction, which assumes no motion occurs within that time frame, we discarded the 4\,s blocks immediately before and after large jumps in the motion estimates. To this end, we calculated a \textit{motion score} between two time points $t$ and $\tau$ as: 
\begin{equation}
    M_{t,\tau} = r_{t,\tau} + d_{t,\tau}
    \label{eq:mscore}
\end{equation} 
with
\begin{equation}
    r_{t,\tau} = R\sqrt{(1-\cos(\mid\theta_{t,\tau}\mid)^2 + \sin(\mid\theta_{t,\tau}\mid)^2}
    \label{eq:mscore_r}
\end{equation}
\begin{equation}
    d_{t,\tau} = \sqrt{(x_{t}-x_{\tau})^2 + (y_{t}-y_{\tau})^2 + (z_{t}-z_{\tau})^2}
    \label{eq:mscore_d}
\end{equation}
where $r$ is the spherical distance calculated on a sphere with radius $R=64\,mm$, $\theta_{t,\tau}$ is the angle of rotation extracted from the Euler angles of the estimated rotations, and \(x_t, y_t, z_t\) and \(x_\tau, y_\tau, z_\tau\) are the positions of the object at times \(t\) and \(\tau\).\cite{tisdall2012volumetric,marchetto2023robust} If the motion score was larger than a threshold, data from both neighboring time points were discarded in the final image reconstruction. The threshold was heuristically set at 1.5\,mm, which visually resulted in the best image quality considering the trade-off between motion and undersampling artifacts. Supporting Fig.\,S2 provides an overview of the amount of data removed across the datasets.

Based on the motion-corrected k-space trajectory and data, the final image reconstruction was performed, also using subspace modeling.\cite{Tamir2017,zhao2018improved,Asslander2018} Here, we used a subspace optimized for the conservation of the Cramér–Rao bound,\cite{mao2024cramer} and we solved the reconstruction problem, which includes a locally low-rank penalty, with the OptISTA algorithm.\citep{jang2023computer} Finally, we estimated parametric maps with a neural network-based method.\cite{zhang2022cramer,cohen2018mr, mao2024bias}

\subsection{Motion Simulations} \label{sec33}
To evaluate the proposed approach, we performed motion simulations by corrupting a reference dataset and performing motion correction on the corrupted dataset. By comparing motion estimates with the ground truth motion, we evaluated the accuracy of motion estimates.

\subsubsection{Reference Dataset} \label{sec331}
As reference dataset, we selected the participant with the least amount of motion ($\bar{M} = 0.22$\,mm) from our pool of 86 acquisitions, where
\begin{equation}
    \bar{M} = \frac{1}{N(N-1)/2} \sum_{t=1}^{N-1} \sum_{\tau=t+1}^{N} \mathbf{M}_{t,\tau}
    \label{eq:pwmscore}
\end{equation}
is the mean \textit{pair-wise} motion score. It can be viewed as a surrogate for the overall data inconsistency, rather than an average of the motion between neighboring time points, as proposed in Ref.~\citen{tisdall2012volumetric}. Here, we used preliminary motion estimates, derived as described in Sec.~\ref{sec32} using a regularization factor $\lambda = 10^{-2}$.

% Figure 3 - qMT maps simulations
\begin{figure}[htb!]
    \centering
    \includegraphics[width=0.95\linewidth]{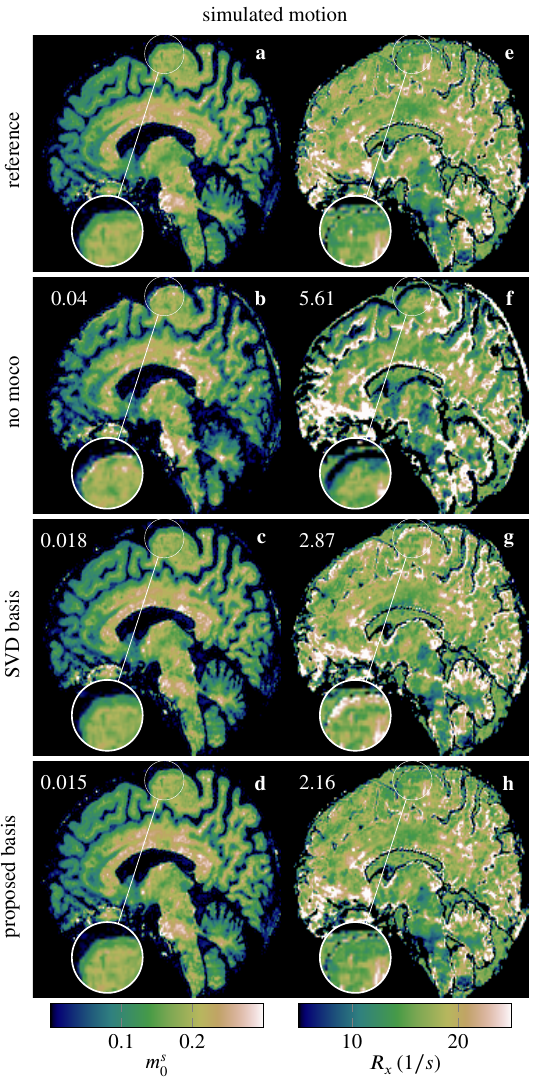}
    \vspace{-10pt}
    \caption{Sagittal view of $m_0^s$ and $R_x$ maps of the selected reference image (no motion added; Sec.~\ref{sec331}) and with simulated motion (b, f), which show strong degradations. In contrast, both the SVD basis (c, g) and the proposed contrast-optimized basis (d, h) significantly improve image quality, with the latter visually agreeing most closely with the reference maps (a, e). The RMSE for each parameter map, compared to the reference (no motion added), is shown in the top left of the respective map. The RMSE was calculated for the entire 3D brain volume, defined by the brain mask. The remaining parametric maps are shown in Supporting Fig.\,S5.}
    \label{fig:sim_maps}
\end{figure}

\subsubsection{Ground Truth Motion} \label{sec332}
Ground truth motion parameters were sourced from a publicly available database (\url{http://34.88.15.16/webapps/home/session.html?app=BrainMRIMotionDB}), initiated by the motion correction research community.
We selected seven motion datasets with unintended head motion that were acquired from pediatric patients. The motion parameters were recorded at a sampling rate of 30\,Hz using an optical tracking system, and we interpolated them to our sampling interval of $T_\text{R} = 3.5\;\text{ms}$ using linear B-splines.
We corrupted the reference dataset with each of the seven sets of motion parameters by rotating each radial k-space spoke as specified by the rotation parameters and by adding a linear phase slope to the corresponding data as specified by the translation parameters.

\subsubsection{Motion Estimation}
We estimated the motion parameters as described in Sec.\,\ref{sec32}.
Although we selected the reference image based on its low motion score, it is not entirely motion-free. 
To account for the baseline motion, we estimated the motion parameters from the uncorrupted and the corrupted reference data. We computed the difference by dividing the affine matrices, and we compared this difference to the ground truth motion parameters. 
This pipeline was performed for each of the seven sets of ground truth motion parameters and separately for the SVD and the proposed contrast-optimized basis.

% Figure 4 - motion parameters in-vivo
\begin{figure*}[htb!]
    \centering
    \includegraphics[width=0.95\linewidth]{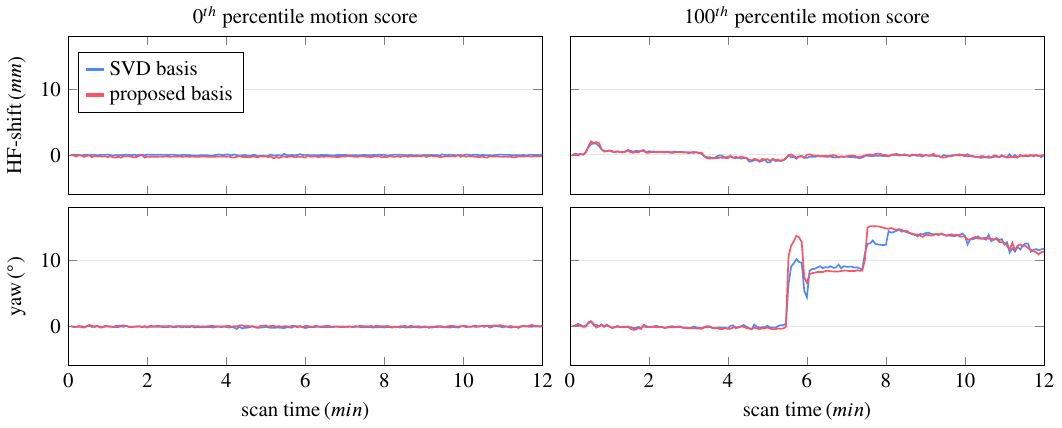}
    \vspace{-10pt}
    \caption{Motion estimates of two extreme acquisitions: the scan with the $0^{th}$ percentile motion score (0.38\,mm and 0.41\,mm with the SVD and contrast-optimized basis; no data removed), and the scan with the $100^{th}$ percentile motion score (10.32\,mm and 10.53\,mm; 16.11\% and 12.22\% of data removed respectively). Here, we show one representative translation (head-foot) and one representative rotation. All 6 motion parameters can be found in the Supporting Fig.\,S2.}
    \label{fig:mopar}
\end{figure*}

\subsubsection{$\lambda$ Optimization} \label{sec333}
The central tuning parameter in our pipeline is the regularization parameter $\lambda$ for the low-resolution reconstruction (see Eq.~\ref{eq:model1}). We used the simulated motion for an objective selection of $\lambda$. As the figure of merit, we defined the \textit{RMSE score} as the $\ell_2$-norm of the root mean squared error (RMSE) between the estimated and ground truth motion parameters:
\begin{equation}
    \lambda_{\text{opt}} = \arg\min_{\lambda} \sqrt{\sum_{i=1}^{3} \text{MSE}_{\text{trans},i}(\lambda) + R^2 \sum_{i=1}^{3} \text{MSE}_{\text{rot},i}(\lambda)}.
    \label{eq:lambda_opt}
\end{equation}
Here, $i$ loops over the three translation and rotation parameters, and the radius $R=64\,\text{mm}$ converts rotations to translations (cf. Eqs.\,\eqref{eq:mscore}---\eqref{eq:mscore_d}).
We determined an optimal $\lambda$ for each of the six flip-angle patterns separately using three of the seven ground truth motion datasets, testing $\lambda \in \{ k e^{-4} \mid k=1,\dots,10 \}$, and choosing the median $\lambda$ between the three datasets (see Supporting Fig.\,S3)

Finally, we evaluated the motion estimation accuracy of the SVD basis and the proposed contrast optimized basis by comparing the RMSE score across the remaining four motion patterns.

\subsection{ROI Analysis} \label{sec34}
To evaluate the performance of the motion correction across our 86 datasets, we performed a region of interest (ROI) analysis. As our pulse sequence uses a radial k-space readout, motion primarily introduces noise-like artifacts. Therefore, we calculated the standard deviation of each quantitative parameter per ROI as a proxy for the artifact level.

We registered the available MP-RAGE to the qMT maps using \textit{Freesurfer}, and segmented the former, focusing on the following ROIs: global white matter, pallidum, corpus callosum, and putamen. The ROI analysis was performed on 85/86 subjects, as one MP-RAGE was excluded due to failed segmentation caused by extensive motion artifacts.

We compared the standard deviation of each qMT parameter and each ROI to the mean \textit{pair-wise} motion score (cf. Sec.~\ref{sec331}). We performed linear regression to determine whether the artifact level increases with motion, hypothesizing that the corresponding slope is reduced when motion correction is performed with the proposed basis.

% Figure 5 - in-vivo qMT maps
\begin{figure*}[htb!]
    \centering
    \includegraphics[width=0.95\linewidth]{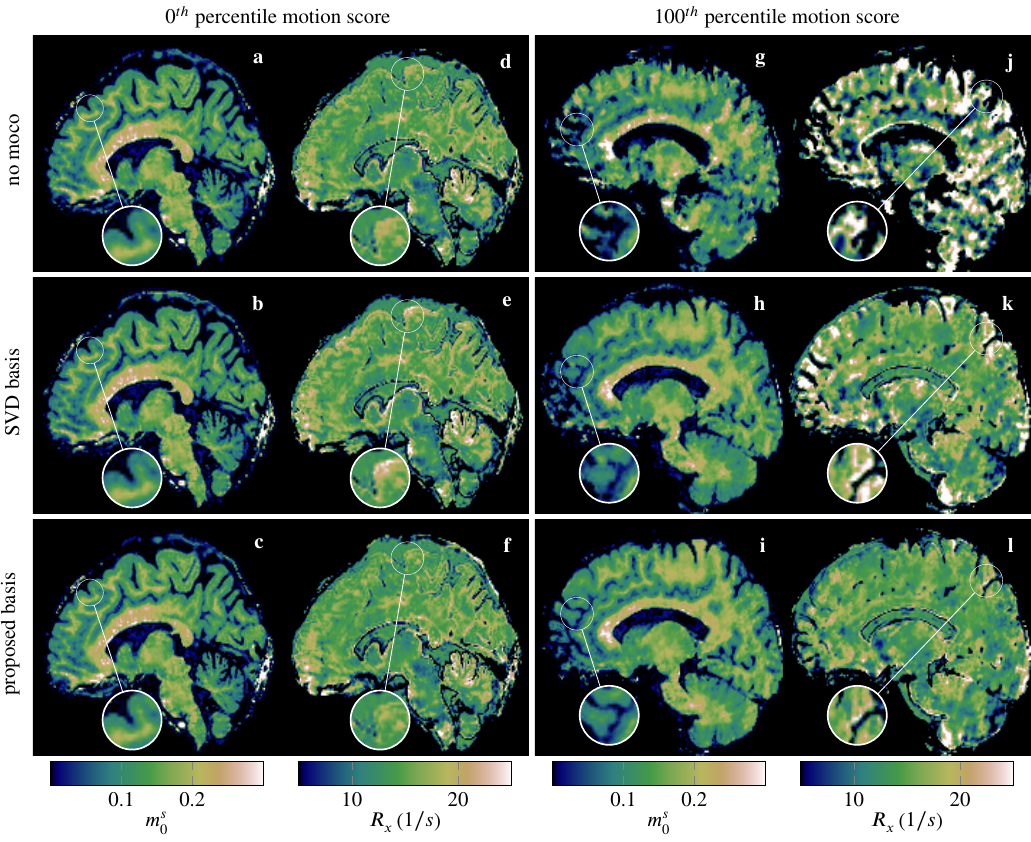}
    \vspace{-10pt}
    \caption{Sagittal view of $m_0^s$ and $R_x$ maps of a scan with minimal motion (a--f). The motion correction with the SVD basis (b,e) introduced subtle artifactual hyperintensities in the $R_x$ map (magnification in e), which are reduced when using the proposed basis (c,f). In the presence of strong motion (100\textsuperscript{th} percentile motion score), the parametric maps reconstructed without motion correction are severely degraded (g,j). Motion correction with the original SVD basis substantially reduced motion artifacts (h,k), which is further improved with the proposed contrast-optimized basis (i,l).
    Like in Fig.~\ref{fig:sim_maps}a, the $m_0^s$ map is expected to have an MP-RAGE-like contrast, while the exchange rate $R_x$ is expected to be near-isointense between gray and white matter, like in Fig.~\ref{fig:sim_maps}e.\cite{asslander2024unconstrained}} 
    \label{fig:volmaps}
\end{figure*}

%% Figure 6 - ROI analysis
\begin{figure*}[htb!]
    \centering
    \includegraphics[width=0.95\linewidth]{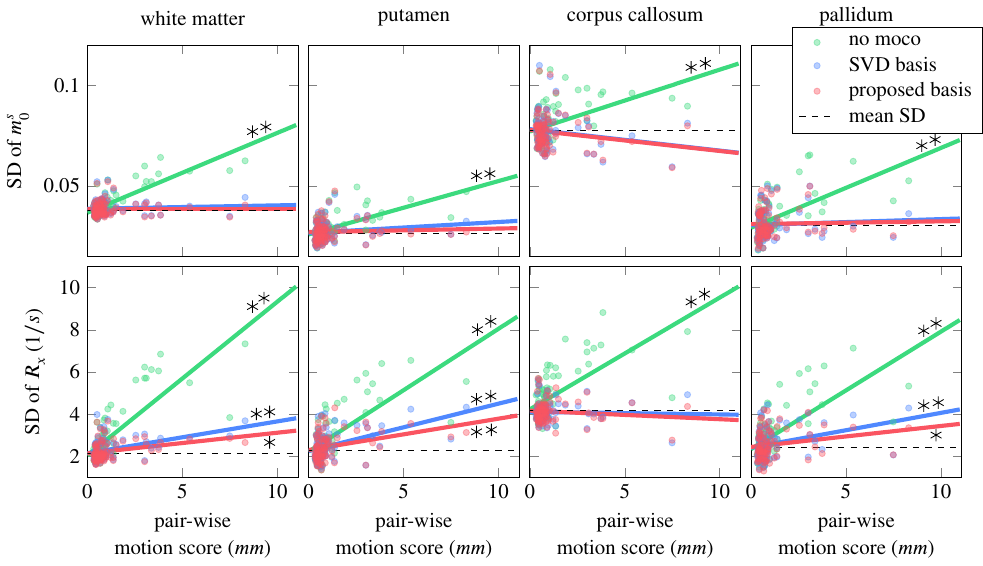}
    \vspace{-10pt}
    \caption{Analysis of noise-like artifacts across all 85 datasets. Each dot represents the standard deviation of the qMT parameter for one participant. The standard deviation was calculated for each motion correction version. Linear regression was performed to analyze the increase in the standard deviation with increasing motion. Black stars denote slopes that differ significantly from zero (* for p-value\,\textless\,0.05, ** for p-value\,\textless\,0.01). The slopes of this linear regression analysis are further analyzed in Fig.~\ref{fig:slopes}.}
    \label{fig:roianalysis}
\end{figure*}

\section{Results}\label{sec4}
\subsection{Simulated Motion} \label{sec41}
Fig.~\ref{fig:mopar_simu} compares motion estimates, using the SVD and the proposed contrast-optimized basis, to the ground truth. Due to the limited temporal resolution, both motion estimates fail to capture high-frequency components of the ground truth motion. Nonetheless, they approximate the slow components of the motion well. 
By visually comparing the estimates based on the two bases, we observe a better approximation when using the proposed basis, particularly during the rapid changes of motion states (cf. magnifications in Fig.~\ref{fig:mopar_simu}). 
This improvement is corroborated by a lower RMSE score (cf. Eq.\,\ref{eq:lambda_opt}) when using the proposed basis (0.35\,mm) compared to the SVD basis (0.38\,mm). 
We note that the RMSE score captures the rapid oscillations of the ground truth motion, providing a lower bound of 0.3\,mm caused by the estimates' temporal resolution of 4\,s. This bound was calculated by downsampling the ground truth. An additional case is shown in Supplementary Fig.\,S6-S7, where the RMSE score decreased from 1.33\,mm (SVD) to 1.26\,mm (proposed). For the other two motion patterns, the RMSE scores are 0.49/0.48\,mm and 0.75/0.66\,mm (SVD/proposed).

Motion correction with either estimates resulted in a significant recovery of image quality, as shown in the $m_0^s$ and $R_x$ maps in Fig.~\ref{fig:sim_maps}. 
When using the proposed approach as compared to the SVD basis, the improved accuracy of the motion estimates translates to enhanced image quality in the parametric maps (cf. magnifications in Fig.~\ref{fig:sim_maps}). 
The improvements are most apparent in the $R_x$ map (g and h vs. e). Still, the $m_0^s$ map also exhibits a subtle bias to lower values in the gray matter when using the SVD basis (c vs. a) that is reduced when using the proposed approach (d).

All motion parameters and the remaining parametric maps are reported in Supplementary Fig.\,S4-S5. Further, an example with stronger simulated motion (motion score of 6.39\,mm) is provided in Supplementary Figs.\,S6-S7, which resulted in an improvement of the RMSE score from 1.33\,mm to 1.26\,mm using the proposed basis.

\subsection{Inherent Motion} \label{sec41}
Fig.~\ref{fig:mopar} shows estimates of representative motion parameters for two participants. The scan on the left has minimal motion, with translations well below 1\,mm and rotations below 1\textdegree. The motion score of this case was 0.41\,mm when using the contrast-optimized basis, which is the smallest across our 85 acquisitions (0\textsuperscript{th} percentile motion score). 
The corresponding parametric maps (Fig.~\ref{fig:volmaps}), including the $R_1^f$ and $R_1^s$ maps (Supporting Fig.\,S9) exhibit slightly reduced artifacts and enhanced details, which suggests that the contrast-optimized basis enables accurate motion estimates. 
In particular, the $R_x$ map (Fig.~\ref{fig:volmaps}) shows fewer hyperintense regions and appears nearly isointense between gray and white matter, which is the expected contrast \cite{asslander2024unconstrained} (cf. Fig.~\ref{fig:sim_maps}a, e).

The second exemplary case has strong motion (motion score 10.53\,mm, corresponding to the 100\textsuperscript{th} percentile, calculated with the proposed basis) and we observe substantial artifacts when reconstructing without motion correction (cf. right column of Fig.~\ref{fig:mopar} and Supporting Fig.\,S8 for all motion parameters). 
Motion correction with either basis substantially improves image quality.
Comparing the maps reconstructed with either basis, we, once again, see artifactual hyperintensities at the edges when using the SVD basis, which are reduced when using the proposed basis (cf. magnifications in Fig.~\ref{fig:mopar} k vs. l). The remaining qMT parameter maps for both participants can be found in Supporting Figs.\,S9-S10. 

Four additional examples of qMT maps corresponding to different levels of motion ($25^{th}$, $50^{th}$, $75^{th}$, $85^{th}$ and $95^{th}$ percentile) can be found in Supporting Figs.\,S11–S20. 
With motions scores between the extreme cases shown in Fig.~\ref{fig:volmaps}, the resulting parametric maps also show artifact levels in between these extremes. 
Overall, we observe that the proposed method either improves the image quality, or has similar image quality compared to the SVD basis.

To evaluate the performance of the motion correction across all 85 scans in our dataset, we calculated the standard deviation for the ROIs mentioned above. We analyzed them as a function of the respective pair-wise motion score (Fig.~\ref{fig:roianalysis}).
Without motion correction, the standard deviation consistently increases with increasing motion, and the slope of a linear regression model differs from zero at a significance level of 0.01 for the majority of parameters and ROIs (see also Supporting Fig.\,S21). 
When performing motion correction, the motion-induced parameter variability is substantially reduced, with most of the regression slopes being non-significantly different from zero, which indicates effective motion correction with either basis. 

\begin{figure}[htb!]
    % \centering
    \hspace{0.2cm}
    \captionsetup[subfigure]{labelformat=empty} 
    \includegraphics[width=.95\linewidth]{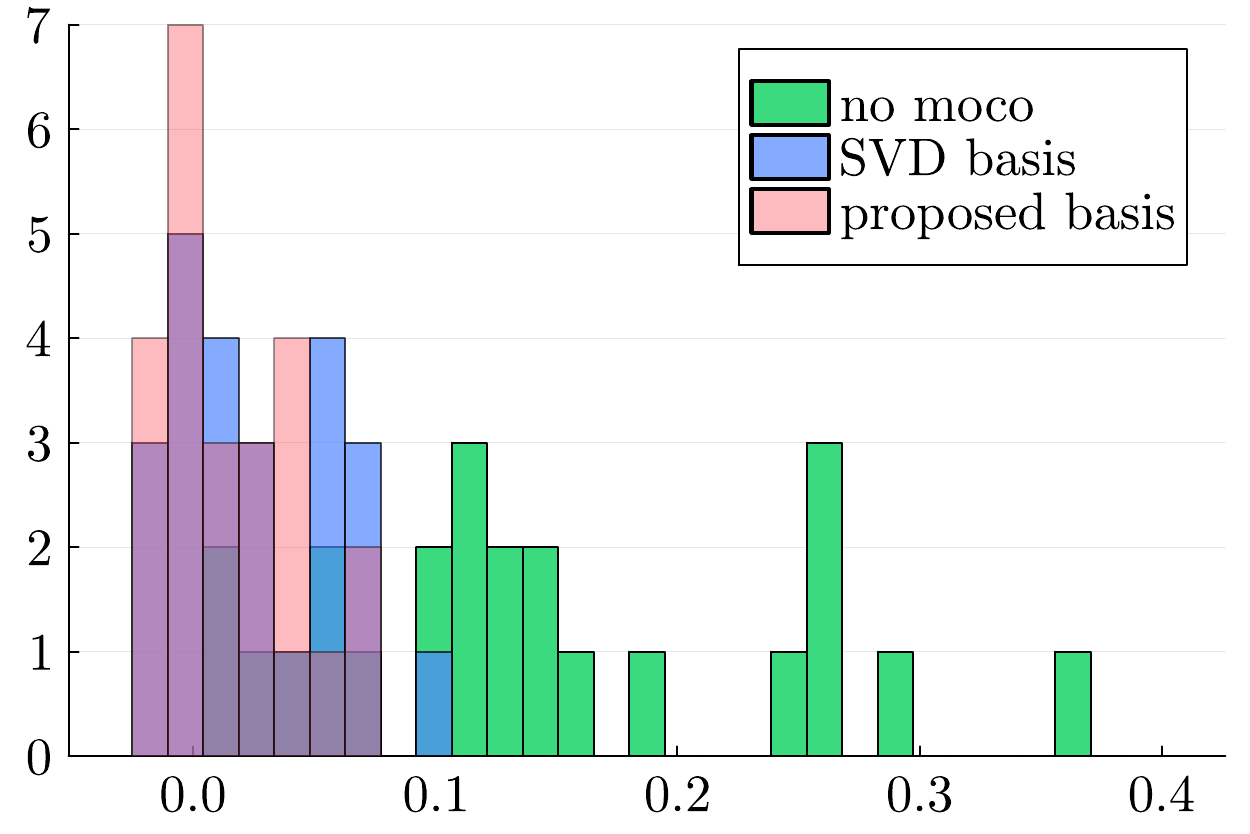}
    \put(-155,-10){\normalsize{slope/intercept (1/$mm$)}}
    \put(-240,65){\rotatebox{90}{\normalsize{Frequency}}}
    \vspace{-0.1cm}
    \caption{
    Analysis of the parameter estimates' standard deviation in various ROIs as a function of the motion score (cf. Fig.~\ref{fig:roianalysis}). The histogram shows the slope of a linear regression, normalized by the respective intercept. This analysis pools all 6 qMT parameters and 4 ROIs. The proposed contrast-optimized basis results in smaller slopes compared to the SVD basis, indicating better motion correction.}
    \label{fig:slopes}
\end{figure}

Based on visual inspection of Fig.~\ref{fig:roianalysis}, the slope of the linear regression appears to be smaller when using the proposed basis in most ROIs and parameters, suggesting improved motion correction. To further evaluate this, we normalized the slope of each parameter and ROI by the respective intercept (no motion). We compared the three reconstructions, pooled over all qMT parameters (6) and ROIs (4) (Fig.~\ref{fig:slopes}). A paired t-test revealed a significant reduction in the normalized slope ($p<0.01$) when using the proposed basis compared to the SVD basis.

\section{Discussion}\label{sec5} 
We proposed enhancing the contrast-to-noise ratio between brain parenchyma and CSF by rotating the SVD basis. 
To this end, we used a generalized eigendecomposition, which is inspired by Region-Optimized Virtual Coils (ROVir).\cite{kim2021region} We demonstrated that the increased contrast overall provides improved accuracy in the motion estimates compared to an SVD basis, generally resulting in better image quality in the parametric maps.

To quantify the improvement in motion estimation accuracy provided by the proposed contrast-optimized basis, we performed simulations applying motion parameters to reference data with virtually no motion. We observed a decrease in the RMSE score when using the proposed contrast-optimized basis compared to the SVD basis in all four motion patterns.

% Figure 7 - slopes analysis

The proposed basis can be used as a one-to-one replacement for a traditional SVD basis. 
Therefore, both approaches have the same target applications, foremost transient-state quantitative MRI techniques, such as MR-Fingerprinting, where the same spin dynamics are repeated while filling the k-space. 
Furthermore, the proposed approach incurs no additional computational costs during reconstruction. Although we tested our method only on brain images and limited our investigation to rigid motion correction, the contrast-optimized basis can also be generated for other body parts with two distinct tissue types. 

Like with the original SVD approach, key factors for successful implementation include 3D imaging to mitigate through-slice motion and a k-space trajectory that provides adequate coverage in each repetition of the spin dynamics, facilitating a time-segmented reconstruction. In this study, we used a koosh-ball trajectory with golden-angle increments, which repeatedly samples the center of k-space. We paired it with a TV-regularized low-resolution reconstruction to mitigate undersampling artifacts. 

While the proposed motion correction can substantially reduce the artifacts in the parametric maps, Fig.~\ref{fig:volmaps} suggests that, in cases of severe motion, the image quality is still impaired despite motion correction with either basis. This is more evident in Fig.~\ref{fig:volmaps} and Supporting Fig.\,S10, which showed the largest amount of motion across all datasets (motion score of 10.53\,mm): despite the considerable improvements obtained after motion correction (especially in the $R_1^f$ maps), the quantitative maps are still severely affected by motion artifacts.
One explanation might be the inherently low temporal resolution of self-navigated motion correction. In our current implementation, each motion state is assigned every 4\,s block, and improving the temporal resolution will be part of future work.
Another source might be coil sensitivities, which are not adapted during the motion-corrected image reconstruction.

The low temporal resolution entails the assumption that no motion occurs during a 4\,s block. To address this limitation, we discard data before and after strong motion. However, in cases of continued strong motion (i.e., substantial variation of the motion states in each 4\,s window), the trade-off between motion and undersampling artifacts imposes a ceiling on the motion correction performance. 

\section{Conclusions}\label{sec6}
We propose a contrast-optimized basis function for self-navigated motion correction in quantitative MR. We utilize the generalized eigendecomposition to enhance the contrast-to-noise ratio between brain tissues and CSF, which improves the accuracy of motion estimates and, ultimately, the image quality of the quantitative parametric maps. The proposed technique does not require any sequence modifications or additional scan time. Consequently, it can be seamlessly integrated into various quantitative MRI methods, e.g., inversion recovery or multi-echo spin echo, where signal variations over time can be effectively captured in a low-rank subspace.

% \section*{Acknowledgments}
% This work was performed under the rubric of the Center for Advanced Imaging Innovation and Research (CAI2R, www.cai2r.net), an NIBIB National Center for Biomedical Imaging and Bioengineering (NIH P41 EB017183).

\subsection*{Author contributions}
Elisa Marchetto: Conceptualization; Data curation; Formal analysis; Investigation; Methodology; Software; Project administration; Writing—original draft; and Writing—review \& editing. Sebastian Flassbeck: Conceptualization; Data curation; Formal analysis; Investigation; Methodology; Software; and Writing—review \& editing. Andrew Mao: Methodology; and Writing—review \& editing. Jakob Assländer: Conceptualization; Investigation; Software; Supervision; Funding acquisition; and Writing—review \& editing. 

% \subsection*{Financial Disclosure}
% The authors have no financial disclosure to report.

\subsection*{Conflict of Interest}
The authors declare no potential conflict of interest.

\subsection*{Data Availability Statement}
The reconstruction pipeline described in Sec.~\ref{sec32} is implemented in \textit{Julia v1.11}, except for the volume registration (as part of the motion estimation) which is performed using SPM12 (Statistical Parametric Mapping) in MATLAB. The codes to generate the proposed contrast-optimized basis are available on \url{https://github.com/EliMarchetto/ContrastOptimized_basis}. 
Unfortunately, we cannot share the raw data, since facial recognition software could be used to identify the participants. 

\bibliography{Bibliography}

\section*{Supporting Information}
Additional supporting information may be found in the online version of the article at the publisher’s website.

\textbf{Figure S1}: In our acquisition scheme, we use six variable flip-angle patterns to encode six biophysical magnetization transfer (MT) parameters. Here we show the corresponding coefficient image derived using the SVD basis (a-f) and the proposed contrast-optimized basis (g-l). 
We observe considerable variability in the contrast between the six flip angle patterns, which can be attributed to a lack of contrast consideration in the optimization objective. In contrast, the proposed basis directly maximizes the contrast between tissues (in this case, brain parenchyma and CSF), resulting in less contrast variability.
The images are reconstructed by aggregating all the radial spokes acquired during one 4\,s RF cycle. More details on the reconstruction in Sec. 3.2.

\textbf{Figure S2}: Distribution of the amount of data removal over the 86 datasets. The data removal is based on a motion score \cite{tisdall2012volumetric} and was calculated separately for the motion estimates derived using the SVD basis \cite{kurzawski2020retrospective} and the proposed contrast-optimized basis.

\textbf{Figure S3}: RMSE scores for simulated motion, calculated by comparing motion estimates to the ground truth. For both the SVD basis and the proposed contrast-optimized basis, we tested ten different regularization parameters $\lambda$ to identify the optimal $\lambda$ (star markers) for each flip-angle pattern. The proposed basis yields lower minimum RMSE scores for most flip-angle patterns across three motion patterns. 

\textbf{Figure S4}: Estimates of moderate simulated motion (motion score of 1.77\,mm). The ground truth was obtained from an online database. The RMSE score, which quantifies deviations of the motion estimates from the ground truth (Sec.\,3.3.3) is 0.38\,mm when using the SVD basis and 0.35\,mm when using the proposed basis. Downsampling the ground-truth motion to 4\,s windows provides a lower bound of 0.3\,mm for the RMSE score, given the low temporal resolution.

\textbf{Figure S5}: Parameter maps in the presence of moderate simulated motion, complementing the $m_0^s$ and $R_x$ maps in Fig.\,3 of the main manuscript. The underlying data was corrupted with the pattern shown in Fig.\,2 and Supporting Fig.\,S4.
The RMSE for each parameter map, compared to the reference (no motion added), is shown in the top left of the respective map. The RMSE was calculated for the entire 3D brain volume, defined by the brain mask.

\textbf{Figure S6}: Estimates of strong simulated motion (motion score of 6.39\,mm). The ground truth was obtained from an online database. The RMSE score, which quantifies deviations of the motion estimates from the ground truth (Sec.\,3.3.3) is 1.33\,mm when using the SVD basis and 1.26\,mm when using the proposed basis. Downsampling the ground-truth motion to 4\,s windows provides a lower bound of 1\,mm for the RMSE score, given the low temporal resolution.

\textbf{Figure S7}: Parameter maps in the presence of strong simulated motion. The underlying data was corrupted with the pattern shown in Fig.\,S6.
The RMSE for each parameter map, compared to the reference (no motion added), is shown in the top left of the respective map. The RMSE was calculated for the entire 3D brain volume, defined by the brain mask.

\textbf{Figure S8}: Estimates of inherent motion using the SVD basis \cite{kurzawski2020retrospective} and the proposed contrast-optimized basis. We compare here a case with minimal ($0^{th}$ percentile motion score) and very strong ($100^{th}$ percentile motion score) unintended motion. The corresponding parametric maps can be found in Fig.\,5 and Supporting Figs.\,S9 and S10.

\textbf{Figure S9}: Parameter maps in the presence of minimal inherent motion (0\textsuperscript{th} percentile motion score), complementing the $m_0^s$ and $R_x$ maps in Fig.\,5 of the main manuscript. In this case, all three reconstructions have virtually identical results.

\textbf{Figure S10}: Parameter maps in the presence of very strong inherent motion (100\textsuperscript{th} percentile motion score), complementing the $m_0^s$ and $R_x$ maps in Fig.\,5 of the main manuscript.

\textbf{Figure S11}: Estimates of inherent motion using the SVD basis \cite{kurzawski2020retrospective} and the proposed contrast-optimized basis. We show here a case with a small amount of unintended motion (motion score 0.65\,mm, $25^{th}$ percentile; no data removed). The corresponding parametric maps can be found in the Supporting Fig.\,S12.

\textbf{Figure S12}: Parameter maps in the presence of small inherent motion (25\textsuperscript{th} percentile motion score). The corresponding motion parameters can be found in the Supporting Fig.\,S11).

\textbf{Figure S13}: Estimates of inherent motion using the SVD basis \cite{kurzawski2020retrospective} and the proposed contrast-optimized basis. We show here a case with an average amount of unintended motion (motion score 0.91\,mm, $50^{th}$ percentile; 1.11\% of data removed). The corresponding parametric maps can be found in the Supporting Fig.\,S14.

\textbf{Figure S14}: Parameter maps in the presence of average inherent motion (50\textsuperscript{th} percentile motion score). The corresponding motion parameters can be found in the Supporting Fig.\,S13).

\textbf{Figure S15}: Estimates of inherent motion using the SVD basis \cite{kurzawski2020retrospective} and the proposed contrast-optimized basis. We show here a case with a substantial amount of unintended motion (motion score 1.28\,mm, $75^{th}$ percentile; no data removed). The corresponding parametric maps can be found in the Supporting Fig.\,S16.

\textbf{Figure S16}: Parameter maps in the presence of substantial inherent motion (75\textsuperscript{th} percentile motion score). The corresponding motion parameters can be found in the Supporting Fig.\,S15).

\textbf{Figure S17}: Estimates of inherent motion using the SVD basis \cite{kurzawski2020retrospective} and the proposed contrast-optimized basis. We show here a case with a substantial amount of unintended motion (motion score 1.82\,mm, $85^{th}$ percentile; 1.11\% of data removed). The corresponding parametric maps can be found in the Supporting Fig.\,S18.

\textbf{Figure S18}: Parameter maps in the presence of substantial inherent motion (85\textsuperscript{th} percentile motion score). The corresponding motion parameters can be found in the Supporting Fig.\,S17).

\textbf{Figure S19}: Estimates of inherent motion using the SVD basis \cite{kurzawski2020retrospective} and the proposed contrast-optimized basis. We show here a case with an extreme amount of unintended motion (motion score 5.17\,mm, $95^{th}$ percentile; 6.11\% of data removed). The corresponding parametric maps can be found in the Supporting Fig.\,S20.

\textbf{Figure S20}: Parameter maps in the presence of extreme inherent motion (95\textsuperscript{th} percentile motion score). The corresponding motion parameters can be found in the Supporting Fig.\,S19).

\textbf{Figure S21}: Analysis of noise-like artifacts across all 85 datasets, complementing the analysis of $m_0^s$ and $R_x$ in Fig.\,6. Each dot represents the standard deviation of the qMT parameter for one participant. The standard deviation was calculated for each motion correction version methods. Linear regression was performed to analyze the increase in the standard deviation with increasing motion. Black stars denote slopes that differ significantly from zero (* for p-value\,\textless\,0.05, ** for p-value\,\textless\,0.01). The slopes of this linear regression analysis are further analyzed in Fig.\,7.

% \appendix
% \input{SupportingMaterial}

\includepdf[pages=-]{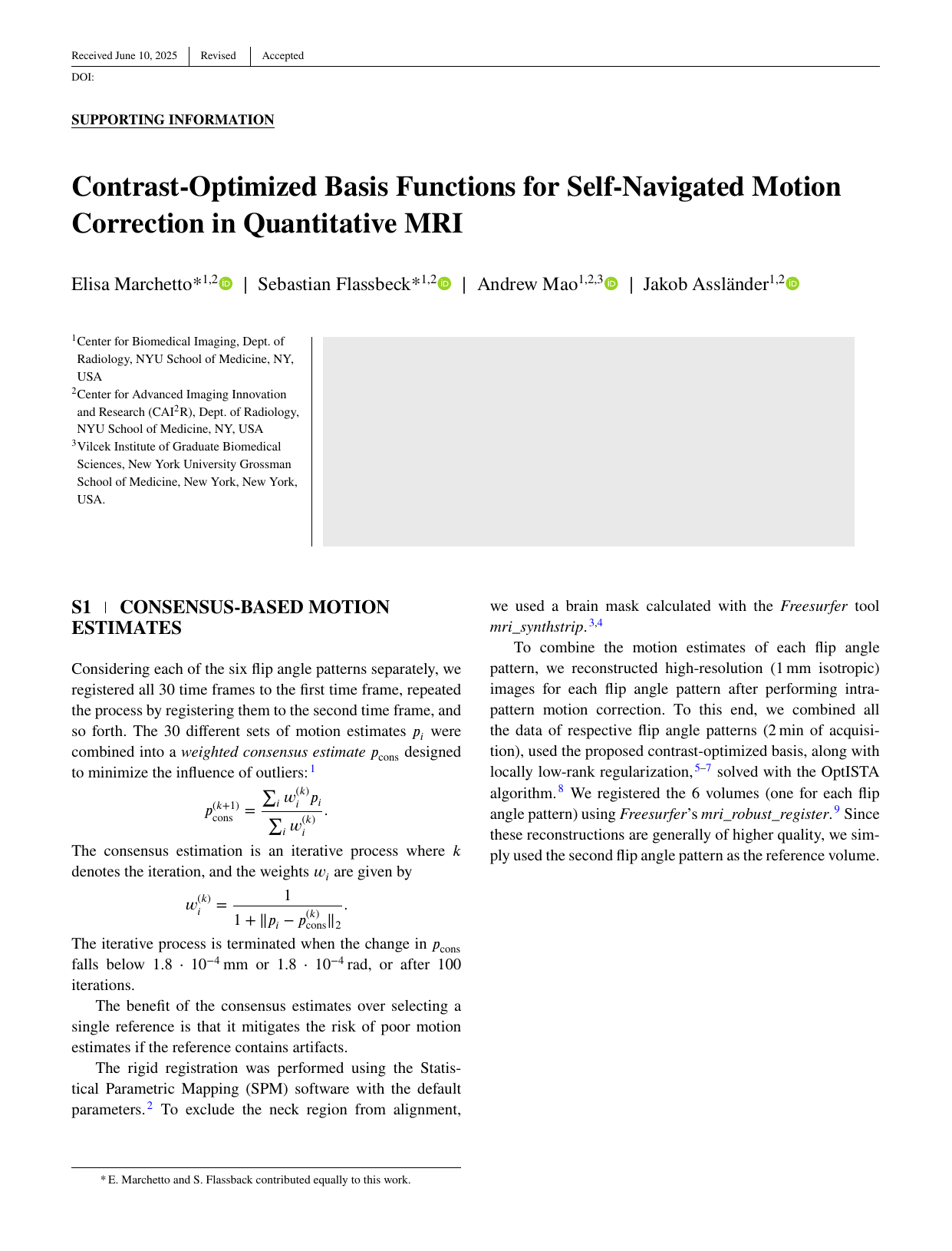} 

\end{document}